\documentclass{elsart}

\usepackage{natbib}
\usepackage{amssymb}

\newcommand{\od}[1]{\mbox{${\mathcal O}(#1)$}}
\newcommand{\BV}{Brunt-V\"ais\"al\"a\ }
\newcommand{\vxi}{\vec{\xi}}
\newcommand{\na}{ \vec{\nabla} }
\newcommand{\lp}{ \left(}
\newcommand{\rp}{ \right)}
\newcommand{\disp}[1]{\displaystyle #1}
\newcommand{\er}{\vec{e}_r}
\def\Div{\mathop{\hbox{div}}\nolimits}
\newcommand{\noi}{ \noindent }
\newcommand{\beq}{\begin{equation}}
\newcommand{\dnr}[1]{\frac{d  #1}{dr}}
\newcommand{\eeqn}[1]{\label{#1}\end{equation}}
\newcommand{\greq}{\begin{equation}\left\{ \begin{array}{l}}
\newcommand{\egreq}{\end{array}\right. \end{equation}}
\newcommand{\ltex}[1]{\quad \hbox{#1} \quad}
\newcommand{\llp}{ \ell(\ell+1)}
\newcommand{\egreqn}[1]{\end{array}\right. \label{#1}\end{equation}}
\def\Ln{\mathop{\hbox{ln}}\nolimits}
\newcommand{\eq}[1]{(\ref{#1})}
\newcommand{\nngreq}{\[\left\{ \begin{array}{l}}
\newcommand{\dnx}[1]{\frac{d  #1}{dx}}
\newcommand{\enngreq}{\end{array}\right. \]}
\newcommand{\dnz}[1]{\frac{d  #1}{dz}}
\newcommand{\infapp}{\raisebox{-.7ex}{$\stackrel{<}{\sim}$}}

\begin{document}
\begin{frontmatter}

\title{Back to the subseismic approximation for core undertones}

\author{Boris Dintrans}
\address{Nordic Institute for Theoretical Physics, Blegdamsvej 17, DK-2100
Copenhagen, Denmark}
\maketitle

\begin{abstract} 
The problem of the long-period gravity modes of the Earth outer fluid core
(the core undertones) is investigated using either the subseismic or the
anelastic approximation. These two approximations aim at filtering out
the uninteresting short-period seismic or acoustic oscillations while
taking into account the density variations across the core. However,
they differ on the form of the equation of mass conservation since
the density perturbations do not contribute to the mass balance in the
anelastic case. Here we show that these two approximations lead to the
almost same results because of the weakness of the core stratification
due to the convective mixing. The anelastic approximation should be
however preferred because it is simpler, mathematically self-consistent
and is the only one which can be applied to problems with a stronger
stratification such as the ones encountered in radiative zones of stars.
\end{abstract}

\end{frontmatter}

\section{Introduction}

To study the long-periods oscillations of the Earth outer core,
geophysicists commonly use the so-called ``subseismic approximation".
This approximation has been proposed by \citet{Smy81} to reduce the
order of the non-rotating governing equations from fourth to second.
It aims at filtering out the high-frequency seismic P-waves from dynamics
while taking into account the density variations across the core. Using
the classical Boussinesq approximation for this problem is indeed quite
satisfactory since, although the acoustic-like waves are filtered out,
the equilibrium density is assumed to be constant \citep{Spi60}.

A similar approximation, the so-called ``anelastic approximation",
is used in meteorology and astrophysics to describe compressible
convection. This approximation has first been derived by \citet{Bat53} and
\citet{Ogu62} to study the dry convection of the Earth troposphere. Their
aim was to eliminate the sound waves from the hydrodynamic equations
since these rapid waves impose very small time steps in the numerical
integration. In addition, they wanted to take into account the density
variations across the atmosphere so the Boussinesq approximation was
still too restrictive. Quite naturally, this anelastic approximation has
been used in astrophysics to the convective zones of stars \citep[see
e.g.][]{Gil81}. This approximation has also been used to describe the
long-period oscillations propagating in the Earth atmosphere and star
interiors. For example, \citet{Dur89} calculated the low-frequency
oscillations of an isothermal atmosphere using the anelastic equations of
\citet{Ogu62}. In the same way, \citet{Din00} computed the oscillations
of rapidly rotating stars. In order to decrease the size of the numerical
rotating problem, a good solution indeed consists to filter out acoustic
quantities while taking into account the large density variations
encountered in radiative zones of stars.

The subseismic approximation has also been used in the field of stellar
oscillations to describe the long-period gravity modes propagating
in non-rotating stars. \citet{Deb92} studied these modes by using the
same subseismic set of equations as in \citet{Smy81}. More recently,
\citet{Din01} proposed a comparative study of the anelastic and subseismic
approximations applied to this problem (hereafter referred to as Paper
I). There we showed that the subseismic approximation is not in fact a
consistent approximation for the equations of motion. This result has been
asserted by comparing the anelastic and subseismic eigenfrequencies with
the complete ones in the case of the homogeneous star and the polytrope
$n=3$. Analytic solutions have been derived for the homogeneous case
and the anelastic eigenfrequencies appear to be about twenty times
more precise than the subseismic ones. In the same way, the computed
anelastic eigenfrequencies are of about five times more precise than
the subseismic ones in the polytropic case. We therefore concluded on
the better efficiency of the anelastic approximation as far as stellar
low-frequency oscillations are concerned.

The aim of this paper is to compare now both approximations in a
geophysical context, namely that of the long-period oscillations of the
Earth outer core. The subseismic approximation has been already tested in
core dynamics and the resulting errors appear to be always at most 
$\od{1\%}$ \citep{Smy84,Cro91,Cro92}. No comparative study with the
anelastic approximation has been however made for this problem. In
particular, the real benefit of the using of the subseismic approximation
instead of the anelastic one is not clear. In fact, we will show
that both approximations lead to the same results for the core
undertones. It means that, contrary to stars, the inconsistency of
the subseismic approximation does not induce unfortunate consequences
in this case. Using the anelastic approximation is however preferable
because it is self-consistent, simpler and can be successfully extended
to more stratified astrophysical problems.

The plan of the paper is the following: after setting out the basic
equations governing the adiabatic oscillations of a non-rotating
fluid (Section \ref{sec1}), we recall the main features of the subseismic
and anelastic approximations (Section \ref{sec2}) and give some details
on the numerical method used to solve the oscillation equations (Section
\ref{sec3}). The results of our calculations are then presented with
the comparisons of the subseismic and anelastic eigenperiods with the
complete ones (Section \ref{sec4}). We conclude with some outlooks of
our results, with in particular their immediate implication to the more
difficult rotating problem (Section \ref{sec5}).

\section{The formalism}

\subsection{The equilibrium model and the complete oscillation equations}
\label{sec1}

The identification by seismology of stable stratification zones in the
Earth outer fluid core is still today controversial. Two stable stratified
zones are in fact expected: {\em (i)} one zone may be located at its
bottom (ICB) due to the release of light elements by the crystallization
of the inner core \citep{Sou91}; {\em (ii)} a second zone may be
located at its top (CMB) due to a thermal buoyancy flux coming from the
mantle \citep{Lis98}. From observations of normal mode eigenperiods,
\citet{Mas79} constrained the minimum value of the core \BV period to be
6h, a value consistent with the core mixing due to convection 
\citep{Cro84,Smy84}. Thus many authors \citep{Cro80,Cro91,Cro92} chose
to take a uniform \BV period of 6h in their studies of core dynamics
and this model has also been used here.

Assuming a time-dependence of the form $\exp(i\sigma t)$ and neglecting
rotation, the adiabatic oscillations of a self-gravitating fluid around
its reference state obey the following linearized equations

\begin{eqnarray}
& & \disp \sigma^2 \vxi = \na \lp \frac{P'}{\rho_0} + \phi'\rp + N^2 \lp
\xi_r - \frac{P'}{\rho_0 g_0} \rp \er, \label{eq1} \\ \nonumber \\
& & \disp \frac{P'}{\rho_0} + c^2_0 \Div \vxi - g_0 \xi_r = 0, \\
\nonumber \\
& & \disp \Delta \phi' = 4\pi G \rho_0 \lp \frac{P'}{\rho_0 c^2_0} +
\frac{N^2}{g_0} \xi_r \rp. \label{eq2}
\end{eqnarray}

\noi where the variables $P'$ and $\phi'$ are respectively the Eulerian
perturbations of pressure and gravitational potential and $\vxi$ is the
Lagrangian displacement. The radial profiles $\rho_0,g_0$ and $c^2_0$
respectively denotes the density, gravity and square of seismic P-wave
velocity and are given by the stable 6h core model of \citet{Cro92} with

\beq 
N^2 = - g_0 \lp \frac{g_0}{c^2_0} + \frac{1}{\rho_0} \dnr{\rho_0}
\rp = \lp \frac{2\pi}{6\hbox{ hr}} \rp^2 = 8.461 \times 10^{-8} \hbox{
s}^{-2}.
\eeqn{bv2}

As in \citet{Cro80}, we assume rigid boundaries both at the ICB ($r=a$)
and CMB ($r=b$) since eigenfrequencies computed using such conditions
are in excellent agreement with those computed for the whole Earth
\citep[see also][]{Cro91}. In addition, the gravitational potential
should be continuous across these surfaces which leads, for a mode of
degree $\ell$, to the following boundary conditions

\greq
\disp \xi_r = \dnr{\phi'} - \frac{\ell}{r} \phi' = 0 \ltex{at} r=a, \\ \\
\disp \xi_r = \dnr{\phi'} + \frac{\llp}{r} \phi' = 0 \ltex{at} r=b.
\egreqn{bc}

\subsection{The inconsistency of the subseismic approximation}
\label{sec2}

The anelastic and subseismic approximations are based on the same
assumption which is that the Eulerian fluctuations $P'$ do not contribute
to the Lagrangian ones $\delta P$, that is

\beq
\delta P = P' + \dnr{P_0} \xi_r = P' - \rho_0 g_0 \xi_r \simeq - \rho_0
g_0 \xi_r \ltex{if} \frac{P'}{\rho_0 g_0} \ll \xi_r.
\eeqn{approx}

\noi Quite surprisingly, this common basic assumption does not lead to
the same approximated form of the equation of mass conservation since

\beq
\hbox{Anelastic: }\Div \vxi + \dnr{\Ln \rho_0} \xi_r = 0, \quad
\hbox{Subseismic: } \Div \vxi - \frac{g_0}{c^2_0} \xi_r = 0.
\eeqn{diff}

We showed in I that the subseismic form of this equation is in fact
incompatible with the basic assumption \eq{approx} since the neglect of
$P'$ in $\delta P$ also implies the neglect of $\rho'$ in $\delta \rho$,
i.e.

\[
P' \ll \dnr{P_0} \xi_r \Rightarrow \rho' \ll \dnr{\rho_0} \xi_r
\ltex{or} \delta P \simeq \dnr{P_0} \xi_r \Rightarrow \delta \rho
\simeq \dnr{\rho_0} \xi_r.
\]

\noi As a consequence, the only consistent form of the equation of mass
conservation is the anelastic one according to

\[
\rho' + \dnr{\rho_0} \xi_r + \rho_0 \Div \vxi = 0 \Rightarrow \Div \vxi
+ \dnr{\Ln \rho_0} \xi_r = 0 \ltex{or} \Div (\rho_0 \vxi) = 0.
\]

\noi The density variations must be neglected in the mass balance if
\eq{approx} is assumed.

An other important result derived in I concerns the uncoupling of the
gravity perturbations $\phi'$. Assuming \eq{approx}, we indeed deduce
the following system from Eqs. \eq{eq1}, \eq{diff} and boundary conditions
\eq{bc}

\greq
\disp \sigma^2 \na \times \vxi = \na \times (N^2 \xi_r \er),  \\ \\
\disp \Div \vxi + \dnr{\Ln \rho_0} \xi_r = 0 \ltex{or} \Div \vxi 
- \frac{g_0}{c^2_0} \xi_r = 0, \\ \\
\xi_r (a) = \xi_r (b) = 0.
\egreqn{syst1}

\noi This second-order set of equations valid for $\vxi$ and $\sigma^2$
clearly forms a well-posed eigenvalue problem so that the subseismic
and anelastic eigenfrequencies can be computed without the taking into
account of the $\phi'$ perturbations.

\subsection{Numerics}
\label{sec3}

As mentioned in Section \ref{sec1}, we choose the same equilibrium model
as in \citet{Cro92}, that is a stable core with a \BV period of 6hr.
This constant period gives us a natural time scale for our problem whereas
the CMB radius $b$ will be the length scale. Then the dimensionless
perturbed fields are expanded onto spherical harmonics with, for
example,

\[
\vxi (r,\theta,\varphi) = \sum_{\ell=0}^{+\infty} u_\ell(r) Y_\ell \er +
v_\ell(r) \na Y_\ell, \quad \frac{P'}{\rho_0} (r,\theta,\varphi) = 
\sum_{\ell=0}^{+\infty} p_\ell(r) Y_\ell,
\]

\noi where $Y_\ell (\theta,\varphi)$ denotes the normalized spherical
harmonic of degree $\ell$ with a zero azimuth $m$ \citep[$m=0$ and no
toroidal component needs to be taken into account since rotation is
neglected; see e.g][]{Rie91}. As an example, the dimensionless form of
the approximated set \eq{syst1} reads

\nngreq
\disp \lambda^2 \lp u_\ell - v_\ell - x \dnx{v_\ell} \rp = u_\ell, \\
\\ \disp x \dnx{u_\ell} + ( 2 + \beta x) u_\ell - \llp v_\ell =
0, \qquad \lp \beta = \dnx{\Ln \rho_0} \ltex{or} \beta = - 
\frac{g_0}{c^2_0}\rp, \\ \\
u_\ell (\eta) = u_\ell (1) = 0,
\enngreq

\noi where $\eta = a/b$ and $\lambda^2=\sigma^2/N^2$. This system may
be formally written as

\[
{\mathcal M}_A \overrightarrow{\Psi}_{\ell n} = \lambda^2_{\ell n}
{\mathcal M}_B \overrightarrow{\Psi}_{\ell n} \ltex{with}
\overrightarrow{\Psi}_{\ell n} = \lp \begin{array}{l}
u_{\ell n} \\
v_{\ell n}
\end{array} \rp,
\]

\noi where ${\mathcal M}_A$ and ${\mathcal M}_B$ denote two differential
operators and $\overrightarrow{\Psi}_{\ell n}$ is the eigenvector
associated with the eigenvalue $\lambda^2_{\ell n}$ of order $n$. This
differential eigenvalue problem is discretized on the Gauss-Lobatto grid
associated with Chebyshev's polynomials. Eigenvalues and eigenvectors
are then computed with the iterative Arnoldi-Chebyshev solver already
used to calculate the oscillations of rotating stars \citep[for more
details on numerics see][]{Din99,Din00}.

\section{Results}
\label{sec4}

In order to compare both approximations, we first calculated
the eigenfrequencies of the complete problem consisting of
Eqs. (\ref{eq1}-\ref{eq2}) plus boundary conditions \eq{bc}. Then, we
compared them with their anelastic and subseismic counterparts computed
from \eq{syst1}. Table \ref{tab} shows the obtained periods (in hours)
for some of the twenty first undertones of degree $\ell = 2$, the periods
being deduced from the dimensionless eigenvalues $\lambda^2_{\ell n}$
using $T_{\ell n}(\hbox{hr})=6/\sqrt{\lambda^2_{\ell n}}$.

The first point is that the lower undertone period $T_1 \simeq 9.6705$
hr well agrees with that computed by \citet{Cro92} who found $T_1 \simeq
9.6741$ hr. The slight difference may be explained both by our different
boundary conditions and the eigenvalue solver since these authors did not
assume rigid boundaries and used a shooting method to compute eigenvalues.

The second point is that the anelastic and subseismic approximated
periods are almost the same. This can be easily understood from
the different forms \eq{diff} of the equation of mass conservation
where the $\xi_r$-term is either equal to $(d\Ln \rho_0/dr)\xi_r$ or
$-(g_0/c^2_0)\xi_r$. Because the \BV frequency is weak, these two terms
are almost equal; indeed from \eq{bv2}, we have

\[
\frac{g_0}{c^2_0} + \dnr{\Ln \rho_0} = - \frac{N^2}{g_0} \sim
- \frac{8.5\times 10^{-8} \hbox{ s}^{-2}}{780 \hbox{ cm/s}^2} \sim - 10^{-5} 
\hbox{ km}^{-1}.
\]

\noi The difference in accuracy between both approximations thus depends
on the stratification strength. As the Earth outer core is almost
adiabatically stratified due to convection, the subseismic and anelastic
approximations give the same results, which are very close to the exact
ones as shown by Table 1. As expected, the relative errors decrease
with increasing order of the mode. The accuracy level of both
approximations is remarkable since, except for the first undertone,
relative errors are less than 0.1\% ! In fact, this very good agreement
is still related to the weak stratification. When applying the anelastic
approximation, one should indeed be careful about the required physical
assumptions. \citet{Bat53} and \citet{Ogu62} derived their original
anelastic set of equations using two assumptions:

\begin{table}
\caption{Periods (in hours) of some of the twenty first $\ell = 2$
undertones of the stable 6hr core, with their corresponding anelastic and
subseismic counterparts and the involved errors in percent.}
\bigskip
\begin{tabular}{lccccc}
\hline
$n$ & Complete & Anelastic & Errors (\%) & Subseismic & Errors (\%) \\ 
\hline
1  & 9.6705371 & 9.6302857 & 0.4162 & 9.6328597 & 0.3896 \\
3  & 23.031630 & 23.012466 & 0.0832 & 23.013470 & 0.0788 \\
5  & 37.465939 & 37.453979 & 0.0319 & 37.454590 & 0.0303 \\
10 & 74.139875 & 74.134154 & 0.0077 & 74.134462 & 0.0073 \\
15 & 110.98888 & 110.98481 & 0.0036 & 110.98502 & 0.0035 \\
20 & 147.88150 & 147.87863 & 0.0002 & 147.87878 & 0.0002 \\
\hline
\end{tabular}
\label{tab}
\end{table}

\begin{enumerate}

\item the reference state around which oscillations occur is almost
adiabatically stratified.

\item the time scale of any disturbance is similar to that of gravity
waves.

\end{enumerate}

Among these two basic assumptions, the second one is the easiest to
fulfil. In this paper, we chose the constant \BV period of our equilibrium
model as the time scale. It automatically satisfies the second assumption
since the computation is restricted to disturbances such as $\partial /
\partial t \sim N$.

The first assumption is more severe. Assuming an almost adiabatic state
means that the equilibrium potential temperature $\theta_0$ is also
nearly constant; i.e. $\varepsilon = \delta \theta_0 / \Theta \ll 1$ where
$\delta \theta_0$ is the jump of the equilibrium potential temperature
accross the layer and $\Theta$ is the reference mean value. \citet{Ogu62}
expanded all dependent variables as a power serie of $\varepsilon$
and obtained the anelastic equations by collecting the first-order
terms. The neglected terms are then $\od{\varepsilon^2}$ and give the
accuracy of the development. This small dimensionless parameter
$\varepsilon$ may be related to the \BV frequency by

\[
N^2 = g_0 \dnz{\Ln \theta_0} \sim \frac{g_0}{d} \frac{\delta
\theta_0}{\Theta} \sim \frac{g_0}{d} \varepsilon,
\]

\noi where $d$ denotes the layer thickness. In our case, we have

\[
\varepsilon \sim \frac{d}{g_0} N^2 \sim \frac{2000\hbox{ km}}
{780 \hbox{ cm/s}^2} 8.5 \times 10^{-8} \hbox{ s}^{-2} \sim 2 \%.
\]

\noi Hence the mean error made when using the anelastic approximation
on the core undertones is about $\od{\varepsilon^2} \sim 0.04\%$ which
agrees well with the results from Table \ref{tab}. The long-period
Earth core oscillations can therefore be seen as a good case for
the anelastic approximation and, despite it inconsistency, also
the subseismic approximation since the almost neutral stratification
prevents unfortunate consequences in this case. On the contrary, as the
stratification increases like for instance in stellar radiative zones,
these two approximations disagree more and more and the anelastic results
become more accurate as shown in I.

\section{Conclusion}
\label{sec5}

We have computed the complete, anelastic and subseismic approximated
eigenperiods of a stable 6h Earth's outer core. After setting out the
complete equations governing the adiabatic oscillations of a non-rotating
and self-gravitating fluid, we presented the basic properties of the
anelastic and subseismic approximations.

Both approximations assume that the Eulerian pressure perturbations
$P'$ do not contribute to the Lagrangian ones $\delta P$, that is
only the fluctuations stemming from the equilibrium pressure gradient
are retained. We then recalled that, unlike the anelastic case where
the perturbed density field has no influence on the mass balance, the
subseismic equation of mass conservation is not compatible with this
assumption (this result being demonstrated in I). As a consequence,
the subseismic approximation is not a consistent approximation of the
equations of motion.

We also recalled that the gravity perturbations decouple from the motion
in both cases. It refutes the common belief in the respect of self-gravity
by the subseismic set of equations; i.e. the only difference with the
anelastic approximation lies in the equation of mass conservation. Then,
two main results have been enlightened by our computations:

\begin{itemize}

\item for the Earth's fluid outer core, anelastic and subseismic
eigenperiods are very close to each other. We explained this result by the
very weak stratification of the fluid. The differences between the
anelastic and subseismic periods indeed come from the stratification
strength. Because of convection, the Earth core is almost adiabatically
stratified and both approximations give the same results.

\item the anelastic relative errors are very small ($\infapp 0.1\%$)
for core undertones. It is not obvious at first sight since the
anelastic set of equations has been initially derived for the
atmospheric dry convection. We explained this very good agreement by
recalling the required physical assumptions implied by the use of
the anelastic approximation: {\em (i)} the reference state should
be almost adiabatically stratified; {\em (ii)} the time scale of
any disturbance should be similar to that of gravity waves. The core
undertones clearly fulfil very well these two basic assumptions and the
anelastic eigenperiods are thus accurate. Because of the weakness of the
\BV frequency, the subseismic results are the same as the anelastic ones
and both approximations can be in fact applied to {\em this problem}.
However, it is more natural to employ the anelastic approximation since
it is simpler, mathematically self-consistent and it is the only one
which can be successfully applied to problems with a stronger
stratification such as the long-period oscillations of stars.

\end{itemize}

The above results have straightforward consequences as far as geophysical
applications are concerned. The main application concerns the rotating
problem, that is to calculate the eigenperiods of the rotating outer
fluid core. \citet{Din99} already calculated the gravito-inertial waves
propagating in a rapidly rotating stratified shell which had the same
aspect ratio than the Earth outer core (i.e. $\eta = a/b \simeq 0.35$). To
decrease the size of the numerical problem, this work has been done under
the Boussinesq approximation so that the density variations across the
outer core have been neglected. Hence an improvement of this problem is
now possible by taking the more complete equilibrium model used in this
paper by the means of the anelastic approximation.

\section{Acknowledgements}

Most of the calculations has been carried out at the Theoretical
Astrophysics Center (TAC, Copenhagen) which is gratefully acknowledged.
This work has been supported by the European Commission under Marie-Curie
grant no. HPMF-CT-1999-00411 which is also gratefully acknowledged. I
also thank Annie Souriau and Michel Rieutord for fruitful discussions.


\end{document}